\documentclass[aps,preprint]{revtex4}
\usepackage{graphicx}
\begin{document}

\title{Correlation effects in a simple model of small-world network}

\author{V. Karimipour }
\email{vahid@sharif.edu}
 \affiliation{Department of Physics,
Sharif University of Technology,P.O.Box 11365-9161,Tehran,Iran}

\author{A. Ramezanpour}
\email{ramzanpour@mehr.sharif.edu}
\affiliation{Department of Physics, Sharif University of
Technology,P.O.Box 11365-9161,Tehran,Iran}

\date{\today}

\begin{abstract}
We analyze the effect of correlations in a simple model of small
world network by obtaining exact analytical expressions for the
distribution of shortest paths in the network.  We enter
correlations into a simple model with a distinguished site, by
taking the random connections to this site from an Ising
distribution. Our method shows how the transfer matrix technique
can be used in the new context of small world networks.
\end{abstract}
\pacs{05.40.-a , 05.20.-y , 89.75.HC} \maketitle
\section{Introduction}

\begin{figure}
\includegraphics[width=8cm,angle=270]{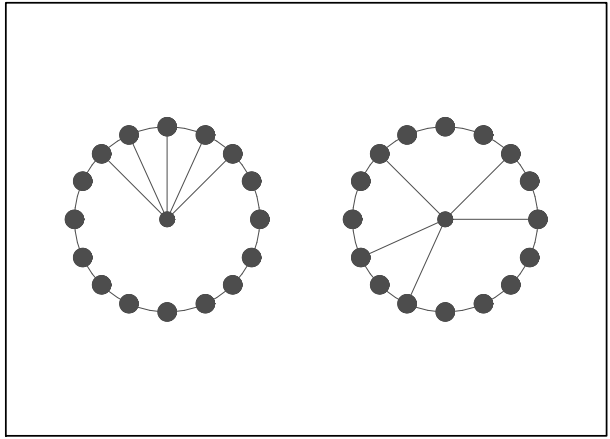}
    \caption{ Two configurations which have different weights in
    our calculations but equal weights in \cite{dm}.
  }\label{figure 1}
\end{figure}

 Real networks, like social networks, neural networks,
power-grids, and documents in World Wide Web
(WWW)\cite{mil,fhcs,n,ajb}, can be modeled neither by totally
random networks nor by regular ones (see \cite{n,w,ab} and
references therein for review). While locally they are clustered
as in regular networks, remote sites have often the chance of
being connected via shortcuts, as in random graphs, hence reducing
the average distance between sites in the network.\\
In a regular networks with $ N $ vertices, the average shortest
path between two vertices $ <l> $ and the clustering coefficient $
C $ scale respectively as $ <l>\sim N $ , and $ C\sim 1 $. The
clustering coefficient $ C $ is defined as the average ratio of
the number of existing connections between neighbors of a vertex
to the total possible connections among them. In random networks,
however we have $ <l>\sim \log N $ and
 $ C\sim 1/ N $ \cite{er,bol}.\\
The properties of many real networks, are a hybrid of these two
extremes, that is in these networks one has $<l>\sim log N$, and $
C\sim 1$. These two effects called collectively small world
effect, are attributed respectively to the presence of shortcuts
and the many inter-connections that usually exist between the
neighboring nodes of such networks \cite{k1,mg,lm}.\\
In 1998, Watts and Strogatz \cite{ws} introduced a simple model of
network showing the small world behavior, which since then has
been investigated as a model of interconnections in many different
contexts, ranging from epidemiology \cite{ak,z,mdl}, to polymer
physics \cite{sc,sab,jb}, and evolution and navigation
\cite{bb1,k2,plh}, The original model of Watts and Strogatz
contained a free parameter $p$, by varying which one could
interpolate between random and irregular networks. Their model,
called hereafter the WS model, consists of a ring of $N$ sites in
which each site is connected to its $2k$ nearest neighbors, hence
making a regular network. After this stage, each bond is re-wired
with probability $p$ to another randomly chosen site. The value of
$p$ tunes the amount of randomness introduced into the network.
Since there is a finite probability of disconnecting the whole
network in this way, Newman and Watts \cite{nw} modified the model
by replacing the re-wiring stage by just addition of shortcuts
between randomly chosen sites on the ring. Since then many more
variants and generalizations of small world networks and their
different characteristics ( e.g. their topology, the properties of
random walks on them, etc.) have been studied. Of particular
interest are three classes of studies. The first class , in which
the static properties  of small world networks have been
investigated \cite{bw,mn,kas,ka},the second class ,where dynamical
aspects have been studied \cite{m,jsb}and the third class,in which
evolving networks are considered  \cite{deb,ke}in order to
generate small world networks  with various connectivity
distributions ,including scale free distributions.\\
In this letter we want to consider another variant of the small
world network, one in which correlation of neighboring nodes in
making connections to remote sites is taken into account(ie;the
presence of a shortcut between two sites affects other shortcuts
in the neighborhood) . For example a node need not make a
short-cut to a remote site if there is such a connection in its
neighborhood. In such networks then, correlations play an
important role. However to perform such a study by exact non-mean
field methods requires a simplification in the original model. We
assume all shortcuts are made via a distinguished site at the
center of the ring. More than being a simplification, this type of
network has practical relevance in many situations where a central
distinguished site governs all the remote interconnections. We
note in passing that such  central sites ,accommodating a large
number of connections,  may exist either in the architecture of
the original networks or else may appear dynamically in evolving
networks \cite{bb2}. In this way we assume that contrary to the
original model,\cite{dm}, the two configurations in
fig.\ref{figure 1} , both with 5 shortcuts are not equiprobable.

\section{The model and Small world quantities}
\begin{figure}
\includegraphics[width=8cm,angle=270]{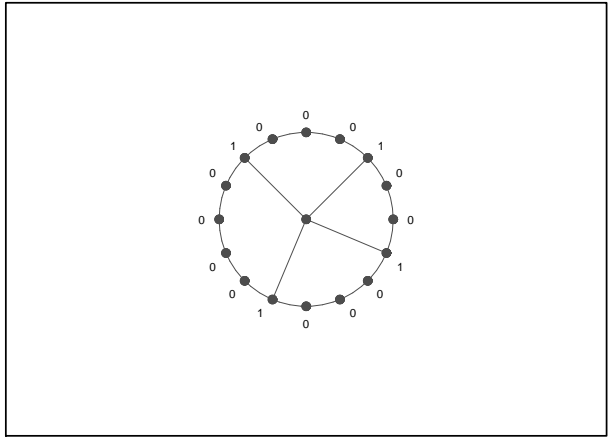}
    \caption{  A simple model of small world networks.
  }\label{figure 2}
\end{figure}

\begin{figure}
\includegraphics[width=8cm,angle=270]{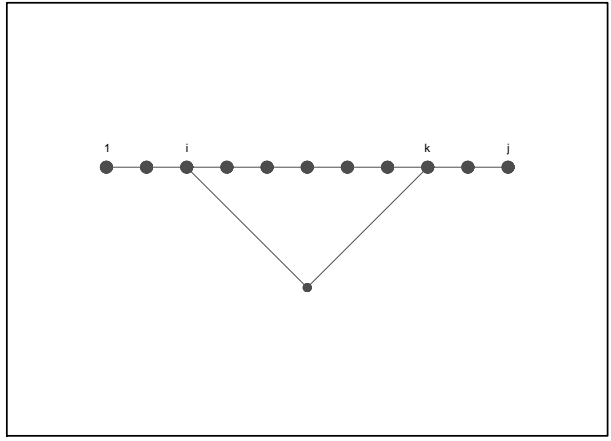}
    \caption{
     A typical configuration which decreases $l_{1j}$.
  }\label{figure 3}
\end{figure}

 We consider a circular network of $N$ vertices with a
distinguished central site \ref{figure 2}. The links on the ring
have unit length. Each shortcut connecting any two sites on the
ring is also of unit length. We assign a random variable $ s_i \in
\{0,1\}$ to each site $i$ of the ring. This random variable is $1$
or $0$ according to whether the site is connected to the center or
not,fig.\ref{figure 2}. Any configuration of these spin variables
corresponds to one and only one configuration of connections to
the center. For example in fig.\ref{figure 1} if each bond is
independently connected to the center with probability $p$, then
the probabilities  of both configurations  are equal and
proportional to $ p^5 (1-p)^{11} $. In general and in the absence
of correlations we will have:
\begin{equation}\label{simple}
  P = \frac{1}{Z}(\frac{p}{1-p})^{s_1 + s_2 + \cdots s_N},
\end{equation}
where Z is a normalization constant. To consider correlations we
generalize the above distribution to an Ising type distribution,
namely to:
\begin{equation}\label{ising} P\{s_i\}=\frac{1}{Z}\biggl({\prod_{i=1}^N
r^{s_{i}} \zeta^{s_i s_{i+1}}} \biggr).
\end{equation}
For $\zeta =1$ and  $r=\frac{p}{1-p}$ we obtain the original
model of \cite{dm}. The value of $\zeta $ controls the
correlations.\\ First let us consider the directed model, i.e; the
links on the circle are directed, say clockwise. Looking at
fig.\ref{figure 3}, we consider a typical configuration like the
one shown in this figure ,in which the nearest shortcuts to sites
$ 1 $ and $j$ are connected at sites $i$ and $k$. This
configuration reduces the distance between sites $1$ and $j$ by an
amount $k-i-1$. Not that the sites between $i$ and $k$ may or may
not be connected to the center. In any such configuration the
quantity $X_{i,k}(1,j)$ defined as:
\begin{equation}\label{xik1}
(1-s_1)\cdots(1-s_{i-1}) s_i s_k (1-s_{k+1})\cdots(1-s_j),
\end{equation}
takes the value $1$.The average of this quantity gives the
probability of such a configuration.In order to find the
probability of the shortest path between sites $1$ and $ j$  to be
equal to $l$, we have to sum over all  those configurations which
give such a shortest path . For $ l \neq j-1 $ the above
probability is given by:
 \begin{equation}\label{dis}
 p(1,j;l)= \sum_{i=1}^{l}< X_{i, j+i-l-1}(1,j)>,
\end{equation}
 where we have used $<..>$ for averaging over
configurations .
Normalization determines $ P(1,j;j-1) $ via:
\begin{equation}\label{p1jj}
 p(1,j;j-1) = 1-\sum_{l=1}^{j-2}p(1,j;l).
\end{equation}
The probability that the shortest path between two arbitrary
vertices be of length $l$, is obtained from:
\begin{equation}\label{pl}
 p(l) = 1/N\sum_{j=l+1}^{N}p(1,j;l).
 \end{equation}
Now the average shortest path between two randomly chosen sites
is:
\begin{equation}\label{lav}
< l> =  \sum_{l=1}^{N-1}lp(l) .
\end{equation}
All the above quantities can be calculated by the transfer matrix
method,in which we write the unnormalized  distribution
\ref{ising} as product of matrix elements  of a matrix $T$ :
\begin{equation}\label{t} T=\left(
\begin{array}
{cc}1&\sqrt{r}\\
\sqrt{r}& r \zeta \end{array} \right)
\end{equation}
with eigenvalues:
\begin{equation}\label{lam}
\lambda_{\pm} = \frac{1}{2}[1+r\zeta \pm \sqrt{(1-r\zeta)^2+4r}].
\end{equation}
The partition function is $Z=\lambda_{+}^N + \lambda_{-}^N$ and
the number of connections per site is given by:
\begin{equation}\label{p}
p := \frac{r}{N}\frac{\partial} {\partial r}\ln Z .
\end{equation}
We now consider the continuum limit of the lattice, where the
number of vertices goes to infinity and the lattice constant goes
to zero as $\frac{1}{N}$ so that the periphery of the lattice is
kept constant at $1$ . We then set: $j/N\longrightarrow x ,
k/N\longrightarrow s, i/N\longrightarrow t$ $l/N\longrightarrow z$
and $ NX_{i,k}(1,j)\longrightarrow X(t,s)(x) $ where the explicit
form of the function $X(t,s)(x)$ will be determined later. We will
then have $ 0\leq x, t, s, z\leq 1 $. Here $x$ is the distance
along the ring.\\
Furthermore we take $ Np(1,j,l)\rightarrow{\cal Q}(x,z) $
,therefore :
 \begin{eqnarray}\label{pxx}
 {\cal Q}(x,z)&=&\int_{0}^{z} {X(t,x+t-z)(x)} \rm{d} t,\cr
 {\cal Q}(z)&=&\int_{z}^{1} {{\cal Q}(x;z)} \rm{d} x,
 \end{eqnarray}
where  ${\cal Q}(x,z)dz$ is the probability that two points whose
distance along the ring is $x$ have a shortest distance between $
z$ and $z+dz$. Then ${\cal Q}(z)dz$ is the probability that the
shortest path between any two points be between $z$ and $z+dz$ .
So $\int_{0}^{1}dz Q(z)=1$  and  finally :
\begin{eqnarray}\label{z}
 < z> &=&\int_{0}^{1} {z {\cal Q} (z)} dz.
 \end{eqnarray}

\subsection{The scaling limit }
Intuitively we expect that in the scaling limit, when $ N
\longrightarrow \infty ,{\rm and }\ \  r = \frac{r_0}{N}$, if we
keep $ \zeta $ finite, then the number of connections to the
center remains finite and in an infinite lattice the
configurations of these connections become quite sparse and hence
correlations can not play a role, at least to leading order. Exact
calculation also verifies this expectation. Here we will consider
a different scaling limit where $ N \longrightarrow \infty ,{\rm
and }\ \  r = \frac{r_0}{N}$ while $\zeta = \zeta_0 N $. This
means that the tendency of an individual one of whose neighbors
has been connected to the center, depends also on the total
population. This assumption is not far from reality, specially in
cases where the center approves a limited amount of connections
and the applicants ,competing for connections, are aware of this
restriction. It turns out that the model shows three distinct
behavior according to the value of the parameter $ r_0 \zeta_0
$.\\
For $r_0\zeta_0 > 1$  we will have:
\begin{eqnarray}\label{lam++}
\lambda_+ &=& r_0 \zeta_0 + \frac{r_0} {N(r_0 \zeta_0-1)},\cr
\lambda_- &=& 1 - \frac{r_0} {N(r_0 \zeta_0-1)},
\end{eqnarray}
and from (\ref{p}) we find:
\begin{equation}\label{pp}
  p = 1 + O(\frac{1}{N}),
\end{equation}
which means that the whole lattice is filled with connections.
Also for $ r_0 \zeta_0 = 1$ we obtain $ p=\frac{1}{2}$,which is
also far from small world regime. To be in the small world regime,
we should keep $ r_0 \zeta_0 < 1$ , which is the case that we will
study in detail.\\
In this case we have:
\begin{eqnarray}\label{lam+}
\lambda_+ &=& 1 + \frac{r_0} {N(1-r_0 \zeta_0)},\cr \lambda_- &=&
r_0 \zeta_0 - \frac{r_0} {N(1-r_0 \zeta_0)}.
\end{eqnarray}
Also from (\ref{p}) we find the total number of connections to be
the finite value:
\begin{equation}\label{M0}
M_0:= Np = \frac{r_0} {(1-r_0 \zeta_0)^2}.
\end{equation}
To calculate $ X_{i,k}(1,j) $ we note that since $ p \rightarrow 0
$ as $ \frac{1}{N} $ , the value of these quantities where either
or both of $ i $ and $ k $ take the extreme values $ 1 $ or $ j $
are suppressed. Using the transfer matrix technique, we obtain
from (\ref{ising} , \ref{xik1}) that for $ 1<i<k<j $,
\begin{equation}\label{xikt}
<X_{i,k}(1,j)>=
T_{00}^{i-2}T_{01}(T^{k-i})_{11}T_{10}T_{00}^{j-k-1}(T^{N-j+1})_{00}
= r (T^{k-i})_{11} (T^{N-j+1})_{00},
\end{equation}
where $ T_{ij}^m = <i|T|j>^{m} $ and $ (T^m)_{ij} = <i|T^m|j> $.
Diagonalizing $T$, using (\ref{t} , \ref{lam}), and taking the
continuum limit, we find after some algebra:
\begin{equation}\label{xst}
X(s-t)(x) = M^2  e^{-M x} e^{M (s-t)},
\end{equation}
where $ M := \frac{r_0} {1-r_0 \zeta_0}$. Inserting this value in
(\ref{pxx}) and integrating we find:
\begin{equation}\label{qxz}
 {\cal Q}(x;z) = z M^2 e^{-M z},\end{equation}
 \begin{equation}\label{qz}
 {\cal Q}(z) =  z(1-z) M^2 e^{-M z}+
 (1+z M)e^{-M z},
 \end{equation}
Turning to  (\ref{z}) it is obtained :
\begin{equation}\label{avz}
< z >= \frac{1}{M}\big(2+e^{-M}\big)-\frac{3}{M^2}
\big(1-e^{-M}\big).
\end{equation}
As expressed in \cite{dm}, these relations already hint at the
emergence of a type of small world behaviour, i.e: with
connecting only 10 sites the average shortest path is reduced
from $\frac{1}{2}$ to 0.17, connecting an extra 10 sites reduces
this value to 0.09.\\
We see that as far as $ \zeta_0 < \frac {1}{r_0}$, the effect of
correlations is only to modify the relations of \cite{dm} by
replacing $ M_0$, the actual number of connections , with an
effective one $ M $. Expressing $ M $ in terms of $ M_0 $ and $
\zeta_0 $ alone, we find: $ M_0 = M (1+M\zeta_0) $, which means
that for low values of $\zeta_0$, $ M \sim M_0 $ while for large
values of $ \zeta_0$ the effective number of connections scales as
the square root of the actual number of shortcuts, $ M \sim
\sqrt{\frac{M_0}{\zeta_0}}$. This effect reflects the tendency of
the shortcuts to get clustered under the influence of
correlations. Hence correlations tend to decrease the small world
effect, since the connections tend to bunch into clusters.

 \section{Undirected and clustered networks}
As far as we have $ N\rightarrow\infty$ and $ M_0=$finite, we can
generalize our results to the cases where

{\bf a)}-the network has no preferred direction

 and

{\bf b)}- each site of the ring is connected to $2k$ of its
neighbors.

In this limit, in going from one site of the ring to another one,
one travels mostly along the ring. Thus denoting the average
shortest paths for the above cases respectively by $\ll z \gg_a$
and $\ll z \gg_b $ we have
\begin{equation}\label{zclus}
 {\ll z \gg}_a = \frac{1}{2}\ll z \gg\hskip.5cm,\hskip .5cm
{\ll z \gg}_b = \frac{1}{k}\ll z \gg ,\end{equation} from which
we obtain
\begin{equation} \label{qaclus} {\cal Q}_{a}(z)
= 2 {\cal Q}(2z)  \hskip 1cm 0\leq z \leq
\frac{1}{2}\end{equation}
 and
 \begin{equation}\label{qbclus} {\cal Q}_{b}(z) = k {\cal Q}(k z)
 \hskip 1cm 0\leq z \leq \frac{1}{k}.
  \end{equation}
And finally ,for the clustered undirected model one will have
 \begin{equation}\label{qab} {\cal Q}_{ab}(z) = 2k {\cal
 Q}(2kz) \hskip 1cm 0\leq z \leq \frac{1}{2k}. \end{equation}

 \section{Conclusion}
We have considered the effect of correlations in a simple model of
small world network, and shown that they generally decrease the
small world effect, since under this condition the connections
tend to bunch into clusters. More concretely in our simple model
the effect of correlations which are controlled by a parameter $
\zeta_0 $, is to reduce (for large $ zeta_0 $) the actual number
of shortcuts $ M_0 $ to an effective one $ M  \sim
\sqrt{\frac{M_0}{\zeta_0}} $, indicating a clustering of
connections to bunches in the lattice.\\Therefore it seems that
the optimal way of designing a small world network would be with
equidistant long-range connections and in order to see the small
world effect and lower the average shortest path, one is better to
use algorithms which anticorrelate the connections.\\ We have
derived our results by exact analytical methods, and have shown
how the transfer matrix technique can be used for obtaining such
properties as average shortest path, or the distribution of
shortest paths in a model of small world network. For all this we
have been forced to study a restricted class of models. No doubt
by doing computer simulations one can study these effects in a
much broader class of models.

\end{document}